\documentclass[12pt]{article}
\usepackage{graphicx}
\usepackage{amssymb}

\large
\newcommand{\be}{\begin{eqnarray}}
\newcommand{\ee}{\end{eqnarray}}
\newcommand{\bea}{\left (\begin{array}{cc}}
\newcommand{\eea}{\right )\end{array}}

\newcommand{\li}[1]{<{#1}>}

\newcommand{\de}{\delta}

\newcommand{\pa}{\partial}
\newcommand{\cd}{{\mathcal D}}
\newcommand{\bcd}{\bar{\mathcal D}}
\newcommand{\na}{\nabla}
\newcommand{\la}{\lambda}
\newcommand{\al}{\alpha}
\newcommand{\ad}{{\dot\alpha}}
\newcommand{\ba}{\beta}
\newcommand{\bd}{{\dot\beta}}
\newcommand{\ofz}{(\zeta)}
\newcommand{\inz}{\!\Big(\frac{-1}\zeta\Big)}
\newcommand{\eg}{{\it e.g.}}
\renewcommand{\u}{{\Upsilon}}
\newcommand{\Om}{{\Omega}}
\renewcommand{\P}{{\Phi}}
\newcommand{\tP}{{\tilde\Phi}}
\newcommand{\mO}{{\mathcal{O}}}

\newcommand{\mat}{\left ( \begin{array}{cc}}
\newcommand{\emat}{\end{array} \right )}
\newcommand{\matt}{\left ( \begin{array}{ccc}}
\newcommand{\ematt}{\end{array} \right )}
\newcommand{\matf}{\left ( \begin{array}{cccc}}
\newcommand{\ematf}{\end{array} \right )}
\newcommand{\vect}{\left ( \begin{array}{c}}
\newcommand{\evect}{\end{array} \right )}
\newcommand{\Tr}{{\rm Tr}}

\begin{document}

\setlength{\baselineskip}{17pt}
\pagestyle{empty}
\vfill
\eject
\begin{flushright}
YITP-SB-04-28\\
hep-th/0405230
\end{flushright}

\vskip 2.0cm
\centerline{\Large \bf{Quivers, Quotients, and Duality}}
\vskip 1.0cm
\centerline{\bf Daniel Robles-Llana\footnote{daniel@insti.physics.sunysb.edu} and Martin Ro\v cek\footnote{Ro\v cek@insti.physics.sunysb.edu}}
\vskip 0.2cm
\centerline{\it C.N. Yang Institute for Theoretical Physics}
\centerline{\it Stony Brook University}
\centerline{\it Stony Brook, New York 11794}
\centerline{\it USA}
\vskip 1.5cm
\centerline{\bf Abstract}

We give a direct computational proof of $N=2$ Seiberg duality for arbitrary quivers, and find the action on the Fayet-Iliopoulos parameters. We also find a new analogous classical duality for K\"ahler potentials of $N=1$ quivers that generalizes the trivial duality 
$Gr(N,N\!+\!M)\simeq Gr(M,N\!+\!M)$ for Grassmannians.

\vfill

\eject
\pagestyle{plain}

\section{Introduction}
\setcounter{equation}{0}
Seiberg duality was originally formulated in \cite{s} as a low-energy  equivalence between $N=1$ supersymmetric gauge theories: an $N=1$ $SU(N_c)$ gauge theory with $N_f$ ``electric'' flavors transforming in the fundamental representation of the color group and no superpotential flows to the same infrared point as an $SU(N_f-N_c)$ gauge theory with $N_f$ fundamental ``magnetic'' flavors interacting through a superpotential with a meson field in the $\bar N_f\!\times\! N_f$ representation of $SU(N_f)$. 

In \cite{aps} the original $N=1$ Seiberg duality was shown to arise as a consequence of an $N=2$ duality. In that work $N=2$ supersymmetric SQCD was broken to $N=1$ by turning on a bare mass $\mu$ for the adjoint chiral superfield $\P$. By analyzing this breaking in the microscopic theory (and sending $\mu$ to infinity) one recovers $N=1$ pure SQCD. By the $N=1$ nonrenormalization theorems this theory should be equivalent to the effective theory at the root of the baryonic branch.\footnote{The authors of \cite{aps} also consider non-baryonic branches.} By the $N=2$ duality the $D$ and $F$ term constraints of the theory on the baryonic branch of the microscopic theory before the breaking to $N=1$ can be mapped to those of an $SU(N_f - N_c)$ theory with $N_f$ flavors, which is the effective theory at the baryonic branch root. The superpotential breaking does not lift the root of the baryonic branch, and when one performs it one recovers $N=1$ $SU(N_f -N_c)$ SQCD with $N_f$ flavors and some extra gauge singlets (the mesons). 

After the discovery that such theories could be embedded in string theory with the inclusion of $D$-branes, many efforts were made to explain this duality from a string theoretical point of view. The first understanding was given in \cite{egk} using generalized Hanany-Witten constructions. This was followed by a more geometric description \cite{ov} using Picard-Lefshetz monodromy and F-theory. More recently, in \cite{hhy}, using $D$-branes probing abelian orbifold singularities, it was shown that Seiberg duality in certain cases arises as a consequence of a symmetry dubbed Toric Duality. A more general approach to deriving Seiberg duality from string theory is put forward in \cite{ckv}. In these works, Seiberg dual theories are engineered from type $IIB$ strings compactified on noncompact Calabi-Yau manifolds which consist of ADE singularities fibered over a complex plane. In that approach, the gauge theory duality is reduced to Weyl reflections on the simple roots of the ADE singularities, or more generally to mutations of exceptional collections of bundles over the Calabi-Yau's. A related analysis was carried out in \cite{bd}. Also, some works derive Seiberg duality from matrix models (see {\it e.g.} \cite{Feng}).

Our approach to Seiberg duality is simpler but in some ways more restricted. We prove that the $N=2$ duality of \cite{aps} (and \cite{ckv}) can be extended to aribitrary Quiver Theories, and holds at the level of K\"ahler potentials.. We find an equivalence relation between different quiver diagrams that encode the field content of some four dimensional supersymmetric gauge theories. In the $N=2$ case, this implies full quantum equivalence of the theories, and was basically known \cite{ap,ckv}. In the $N=1$ case, we find a classical equivalence relating different gauge theories whose infrared limit has the same K\"ahler potential. Since we prove an exact algebraic equivalence of quotients, we also have shown that the $D=2$ superconformal field theories corresponding to these quivers are the same.

In the $N=2$ case, the quivers encode hyperk\"ahler quotients; we use the projective superspace formalism \cite{pr,ur}, to prove $N=2$ Seiberg duality for general $N=2$ quivers with arbitrary numbers of gauged flavors. We see that this generalization entails a non-trivial mapping among the $FI$ parameters associated to the node on which we perform the duality, and the adjacent ones. Noticing the formal similarity of hyperk\"ahler quotients in the projective superspace formalism with ordinary K\"ahler quotients in $N=1$ superspace, we discover a new kind of classical ``Seiberg duality'' for certain K\"ahler quotients; this generalizes the trivial duality 
$Gr(N,N\!+\!M)\simeq Gr(M,N\!+\!M)$ for Grassmannians to broad classes of quivers.

The paper is organized as follows: in the next section, we give a simple mathematical 
statement of our results along with a few examples, as well as a brief physical description. 
In the next section, we use the language of supersymmetric gauge theories to derive 
our $N=1$ results in $N=1$ superspace. We then derive our $N=2$ results in projective superspace, and then rederive them in $N=1$ superspace; this should be useful for 
studying the case where we break $N=2$ supersymmetry with a superpotential. 
Finally, in the appendix, we present a simple proof that the ADE quivers are the 
unique ones that are self-dual under Seiberg duality.\footnote{We thank Anthony 
Knapp for providing us with the proof.}

\section{Results}
\setcounter{equation}{0}
\subsection{Mathematical description}

We consider $N=1$ and $N=2$ quivers. An $N=1$ quiver is a labeled graph 
with nodes $i$ and directed links $\!\li{\!\!ij\!\!}$ connecting some of the nodes.
To each node we associate a complex vector space $V_i$ of dimension $n_i$, a
unitary group $G_i\equiv U(n_i)$, and a nonvanishing real number $c_i$; to
each link we associate the space $Hom(V_i,V_j)$ of complex linear maps
$\phi_{ij}$ from $V_i$ to $V_j$. Clearly, since at each node there is a natural
action of $G_i$ on $V_i$ (the fundamental representation of $U(n_i)$), a
natural action of the product group $\prod_i G_i$ is induced on the direct sum $\sum_{\li{ij}}Hom(V_i,V_j)$. The space that we study is the K\"ahler 
quotient of this flat complex space by the product group, with the level of the 
moment map\footnote{Explicitly, the moment map constraint at the node $i$ 
with outgoing links $\li{ij}$ and incoming links $\li{ji}$ is given by $\sum_{j\in\{\li{ij}\}} \phi_{ij}\circ\bar\phi_{ji}-\sum_{j\in\{\li{ji}\}} \bar \phi_{ij}\circ\phi_{ji}= c_i 
\mathbb{I}_{n_i\!\times\! n_i}$.} of the $U(1)$ 
factor of each $G_i$ given by $c_i$. As each component $Hom(V_i,V_j)$ of 
the vector space whose quotient we take transforms in the bifundamental 
representation $(n_i,\bar n_j)$, the overall diagonal $U(1)$ subgroup of 
the product group does not act; its corresponding moment map constraint restricts the levels:
\be
\sum_i  n_i\,c_i = 0~;
\label{cc}
\ee
equivalently, we consider the K\"ahler quotient with respect to $S(\prod_i G_i)$. Such
a quotient may be a complete manifold, or it may be a variety with singularities at points 
where the isotropy subgroup of the quotient group changes (if the moment map
constraint does not exclude such points).

An $N=2$ quiver is almost identical, but in addition, each node has associated to it a complex number $b_i$ as well as $c_i$, the links are now bidirectional, that is, each link carries the direct sum $Hom(V_i,V_j) \oplus Hom(V_j,V_i)$, and we consider the {\em hyper}k\"ahler quotient with the levels of the quaternionic $U(1)$ moment maps given by $b_i$ and $c_i$. The constraint (\ref{cc}) now has a counterpart\footnote{In this case, there is a real and a complex moment map constraint at each node: $\sum_{j\in\{\li{ij}\}} \left(\phi_{ij}\circ\bar\phi_{ji}- \bar\phi_{ij}\circ\phi_{ji}\right) =c_i\mathbb{I}_{n_i\!\times\! n_i}$ and $\sum_{j\in\{\li{ij}\}} \phi_{ij}\circ\phi_{ji}=b_i\mathbb{I}_{n_i\!\times\! n_i}$.} 
\be
\sum_i  n_i\,b_i = 0~.
\label{cb}
\ee

We have not sorted out the most general $N=1$ case, but we have found that two different quivers give rise to the {\em same} quotient manifold if we transform any node $i$ that has only incoming or outgoing links (a ``maximally anomalous'' node) by reversing the 
direction of the links, changing 
\be
n_i\to \left(\sum_{j\in\{\li{ij}\}} n_j \right)-n_i~,
\label{nd}
\ee
and mapping the levels 
\be
c_i\to -c_i~~\hbox{and}~~\forall j\in\{\li{ij}\},~ c_j\to c_j+\ell_{\li{ij}}c_i~,
\label{cd}
\ee
where $\ell_{\li{ij}}$ is the number of links between the nodes $i$ and $j$.
Note that (\ref{nd}) and (\ref{cd}) conspire to preserve (\ref{cc}). The levels 
of the moment maps play a crucial role; indeed, in the singular case (level$=0$), 
the duality does {\em not} hold. 

Some simple examples duality are shown in the table below. These are given by quivers with two nodes with dimension $V_{1,2}=(n_1,n_2)$ connected by $\ell$ links all with the same orientation. 
\begin{table}[htbp]
\begin{center}
\begin{tabular}{|c|c|c|c|c|}
\hline
Manifold & dim $V_1= n_1$ & $dim V_2= n_2$ & $\ell$-links & Quotient\\
\hline
 Gr$(k,\ell)$ & 1 & $k$ & $\ell$ & $\mathbb{C}^{k\ell}/\!/U(k)$\\
 ~ & 1 & $\ell\!-\!k$ & $\ell$ & $\mathbb{C}^{(\ell\!-\!k)\ell}/\!/U(\ell\!-\!k)$\\
 ~ & $k\ell\!-\!1$ & $k$ & $\ell$ & $\mathbb{C}^{k(k\ell\!-\!1)\ell}/\!/S[U(k\ell\!-\!1)U(k)]$\\
\hline
$\mathbb{CP}^2$ & 1 & 1 & 3 & $\mathbb{C}^3/\!/U(1)$\\
~ & 1 & 2 & 3 & $\mathbb{C}^6/\!/U(2)$\\
~ & 5 & 2 & 3 & $\mathbb{C}^{30}/\!/S[U(5)U(2)]$\\
~ & 5 & 13 & 3 & $\mathbb{C}^{195}/\!/S[U(5)U(13)]$\\
~ & $F_{2n-1}$ & $F_{2n+1}$ & 3 & $\mathbb{C}^{3F_{2n-1}F_{2n+1}}/\!/S[U(F_{2n-1})U(F_{2n+1})]$\\
\hline
$\mathbb{C}^0$ & 1 & 0 & $\ell$ & $-$\\
~& 1 & $\ell$ & $\ell$ & $\mathbb{C}^{\ell^2}/\!/U(\ell)$ \\
~& $\ell^2-1$ & $\ell$ & $\ell$ & $\mathbb{C}^{\ell^2(\ell^2-1)}/\!/S[U(\ell^2-1)U(\ell)]$ \\
\hline
\end{tabular}
\end{center}
\label{examples}
\caption{Some dual examples}
\end{table}

\noindent For $(n_1,n_2)=(1,k)$, this is the K\"ahler quotient $\mathbb{C}^{k\ell}/\!/U(k)$, which is the complex Grassmannian $Gr(k,\ell)$ of $k$-hyperplanes in $\mathbb{C}^\ell$. Applying the duality at node $2$ gives $(n_1,n_2)=(1,\ell\!-\!k)$, which is just $Gr(\ell\!-\!k,\ell)$ and is well known to be the same as $Gr(k,\ell)$. Applying the duality at node $1$ instead, we get $(n_1,n_2)=(k\ell\!-\!1,k)$, which is the K\"ahler quotient $\mathbb{C}^{k\ell(k\ell\!-\!1)}/\!/S[U(k)U(k\ell\!-\!1)]$, which is certainly not immediately recognizable as $Gr(k,\ell)$; one can continue dualizing the two nodes alternately and get a whole series of K\"ahler quotients that all give rise to the same manifold. For example, for $\ell=3$, we can start with $(n_1,n_2)=(1,1)$, which is just $\mathbb{CP}^2$; applying our duality, we find the sequence $(1,1)\simeq(1,2)\simeq(5,2)\simeq(5,13)\simeq\ldots\simeq(F_{2n-1},F_{2n+1})$, where $F_n$ is the $n'th$ Fibonacci number; thus $\mathbb{CP}^2\simeq \mathbb{C}^{3F_{2n-1}F_{2n+1}}/\!/S[U(F_{2n-1})U(F_{2n-1})]$.  This resembles the results of \cite{f}, but is now applied directly to the actual quotient spaces. 

The last set of examples in the table involve duals of a null node (dim$V_i=0$); null nodes
will be discussed in great detail in \cite{rll}. Further examples involving a null node 
are shown in Fig.~1 below:
\begin{figure}[htbp]
\begin{center}
\includegraphics[width=105mm]{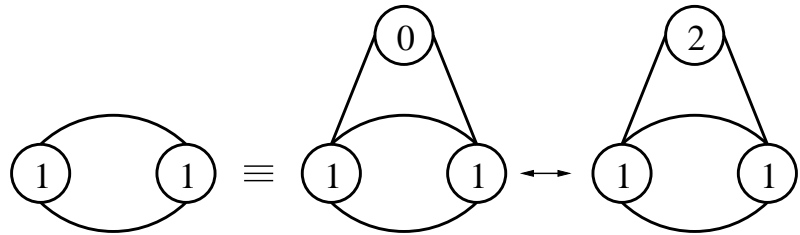}
\caption{Duality involving a null node}
\end{center}
\label{fig1}
\end{figure} 

For the $N=2$ case, since the links are bidirectional, our results apply to all quivers; the only modification is that the complex levels $b_i$ transform in the same way as the real levels $c_i$ in (\ref{cd}). In both cases, the levels give the moduli of the quotient metrics. All the examples can also be considered in the $N=2$ case.

We have also studied which quivers are preserved by the map (\ref{nd}) up to changes in the levels. As shown in the appendix, these are precisely the extended Dynkin diagrams of the ADE series of Lie algebras. A related proof appeared already in \cite{Katz,He}, where it was shown that ADE Dynkin diagrams were the only ones to yield four-dimensional super-conformal theories. In this case, the action of the duality on the levels amounts to an identification in moduli space by Weyl reflections, and thus the moduli spaces of these quotients are wedges \cite{ckv,katzmorrison}.

\subsection{Physical interpretation}

These results have a simple physical description. The quivers encode data for $N=1$ and $N=2$ supersymmetric gauge theories; the gauge group is the product of the unitary groups associated to the nodes, and the matter fields are associated to the links; in the $N=1$ case, the matter field on a link with orientation $i\to j$ is a chiral superfield in the bifundamental $(n_i,\bar n_j)$ representation of the gauge group, whereas in the $N=2$ case, the (unoriented) links represent hypermultiplets in the $(n_i,\bar n_j)\oplus(n_j,\bar n_i)$ representation of the gauge group. The $N=1$ moment map constraints are the $D$-term constraints; the real $N=2$ moment map constraints are $D$-term constraints, whereas the holomorphic moment map constraints are the $F$-term constraints. The real constants $c_i$ are the Fayet-Iliopoulos (FI) parameters in the $D$-term constraints, and the complex constants $b_i$ are the $N=2$ FI parameters in the $F$-term constraints.

\section{$N=1$: The K\"ahler quotient}
\setcounter{equation}{0}
We begin with the simplest case of classical Seiberg-like duality. 

\begin{figure}[htbp]
\begin{center}
\includegraphics[width=105mm]{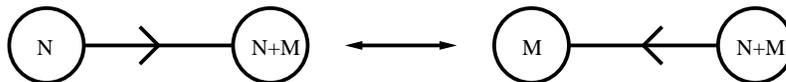}
\caption{Dualizing one node of a simple quiver.}
\end{center}
\label{fig2}
\end{figure}
\noindent Consider a quiver (left side of Fig.~2) with two nodes corresponding 
to gauge groups $U(N)$ and $U(N\!+\!M)$ (with $N,M{\geq}0$), and a single 
link representing a complex (chiral) superfield $\P$ in the 
$(N,{\overline{N\!+\!M}})$ representations of the gauge groups at the nodes. 
This is the field content necessary to perform the K\"ahler quotient of ${\mathbb{C}}^{N(N\!+\!M)}$ by the group $U(N)\!\times\! U(N\!+\!M)$.  More generally, we can see this quiver as a part of a bigger diagram; we are however interested in dualizing node $N$ only, and nodes that are not connected to it will play no role in the discussion.  As is well known, to carry out the K\"ahler quotient, one has to write the real moment map equations ($D$-terms) for the holomorphic action of the gauge group on ${\mathbb{C}}^{N(N\!+\!M)}$. To find the quotient space, one restricts oneself to a given level set of the moment maps and divides by the gauge group. Alternatively, the quotient space is found as the set of stable orbits under the complexified gauge group. To actually find the metric on this space, the most convenient way to proceed is to gauge the K\"ahler potential on the covering space by introducing a vector superfield $V$ and couple it to the chiral fields on ${\mathbb{C}}^{N(N\!+\!M)}$. The vector superfield complexifies the gauge group by lifting the gauge symmetry into superspace \cite{hklr}. Solving the equations of motion for $V$ yields the metric on the quotient. 

The gauged K\"ahler potential can be written as
\be
\label{kg}
K=\Tr({\bar\P }e^{V_N}\P e^{-V_{N\!+\!M}})-c_N\Tr\,V_N-c_{N\!+\!M}\Tr\,V_{N\!+\!M}+{\mathcal{L}}(extra)~.
\ee
In the above expression $c_N$ and $c_{N\!+\!M}$ are the $FI$ parameters corresponding to the $U(1)$ factor of $U(N)$ and $U(N\!+\!M)$, respectively, and ${\mathcal{L}}(extra)$ stands for the contributions to the gauged K\"ahler potential coming from the rest of the quiver diagram.

We want to compare this expression to the one arising from the (part of the) quiver on the right of Fig.\ 2. The gauged K\"ahler potential for the second diagram can be written
\be
\label{tkg}
{\tilde{K}}=\Tr ({\tP }e^{-V_{M}}{\bar{\tP}}e^{V_{N\!+\!M}})-{\tilde{c}}_{M}\Tr(V_{M})-{\tilde{c}}_{N\!+\!M}\Tr\,V_{N\!+\!M}+{\mathcal{L}}(extra)~.
\ee
Note the extra minus sign in front of $-V_M$ and $V_{N\!+\!M}$ in the first term; this represents a reversal 
in the orientation of the arrow on the link. Using the equations of motion, we solve for the gauge field at 
the node we are dualizing in the original gauged K\"ahler potential. The equations of motion for $V_N$ 
read
\be
{(e^{V_N})^a}_{b}{(\P e^{-V_{N\!+\!M}}{\bar\P })^b}_c=c_N{{\delta}^a}_c~.
\ee
We solve this equation as
\be
{(e^{-V_N})^a}_{b}=\frac1{c_N}{(M)^a}_{b}~,
\ee
where we have denoted $M\equiv\P e^{-V_{N\!+\!M}}{\bar\P }$. Substituting this back into the gauged K\"ahler potential we obtain
\be
K&=&c_N\Tr \ln(M)-c_{N\!+\!M}\Tr\,V_{N\!+\!M}+{\mathcal{L}}(extra)\nonumber \\ 
&=&c_N \ln \det(M)-c_{N\!+\!M}\Tr\,V_{N\!+\!M}+{\mathcal{L}}(extra)~,
\ee
where we have dropped irrelevant constant terms. Similarly, for the dual quiver we find
\be
~~{\tilde{K}}&=&-{\tilde{c}}_{M} \ln \det(\tilde{M})-{\tilde{c}}_{N\!+\!M}Tr\,V_{N\!+\!M}+{\mathcal{L}}(extra) ~,
\ee
with ${\tilde{M}}\equiv{\bar{\tP }}e^{V_{M+N}}{\tP }$. To prove the equivalence of these two K\"ahler potentials we proceed as follows. Using the $U(N)$ gauge symmetry in the original theory we can bring $\P $ to the form
\be
\label{p}
\P = 
\left( 
\begin{array}{cc} 
\hbox{\bf{1}}_{N\!\times\! N} & \Om_{N\!\times\! M} 
\end{array} 
\right)~.
\ee
Similarly, using the $U(M)$ gauge group in the dual theory we can write
\be
\label{tp}
{\tP} = 
\left( 
\begin{array}{c} 
{\Sigma}_{N\!\times\! M} \\ {\bf{1}}_{M\!\times\! M} 
\end{array} 
\right)~.
\ee
Now define the square matrix
\be
\hat\P =
\left( 
\begin{array}{cc} 
{\bf{1}}_{N\!\times\! N} & \Om_{N\!\times\! M} \\ 
{\bf{0}}_{M\!\times\! N} & {\bf{1}}_{M\!\times\! M} 
\end{array} 
\right)~.
\ee
Note that this matrix has determinant $1$, from which it follows that
\be
\label{dethat}
\det({\hat\P }e^{-V_{N\!+\!M}}{\bar{\hat\P }})=\det(e^{-V_{N\!+\!M}})~.
\ee
However, for {\em any} invertible matrix $\mO= \left( \begin{array}{cc} A & B \\ C & D \end{array} \right)$ with inverse $\mO^{-1}= \left( \begin{array}{cc} P & Q \\ R & S \end{array} \right)$, the following identity holds 
\be
\label{id}
\det\mO= \frac{\det A}{\det S}~.
\ee  
Now consider
\be
\mO=\hat\P e^{-V_{N\!+\!M}}\hat{\bar\P}\equiv
\left(
\begin{array}{cc}
 M   &  * \\
 \, *\,    &  *  \\
\end{array}
\right)~,
\ee
where we do not need the form of the entries indicated with a $*$, and we have used (\ref{p}). We now choose ${\Sigma}=-\Om$. Then
\be
{{\hat\P }}^{-1}
= \left( \begin{array}{cc} {\bf 1} & \Sigma \\ 0 & {\bf 1} \end{array} \right)~,
\ee
and using (\ref{tp}), we find
\be
\mO^{-1}=
\left(
\begin{array}{cc}
 \, *\,    &  * \\
 \, *\,    & \tilde{M}  \\
\end{array}
\right)~.
\ee
Applying the identity (\ref{id}) and using (\ref{dethat}), we obtain
\be
\label{detdet}
\frac{\det M}{\det\tilde M}=e^{-V_{N\!+\!M}}~.
\ee
Plugging this back into the initial gauged K\"ahler potentials, we see that (\ref{kg}) and (\ref{tkg}) identical provided the FI parameters are related as
\be
-c_N&=&{\tilde{c}}_{M}~,\nonumber \\ 
c_{N\!+\!M}+c_N&=&{\tilde{c}}_{N\!+\!M}~.
\ee
We have thus shown that the K\"ahler quotient corresponding to two
quivers related by this classical Seiberg duality is the same. This is an interesting mathematical fact; its physical significance is unclear, as the gauge groups that we quotient by are anomalous.

So far we have restricted our attention to the simplest case, in which the would-be dualized node is connected to a single neighbor. The general case follows easily.  To see that, consider the quiver in Fig.\ 3. The relevant part of the gauged K\"ahler potential is now
\be
\label{kp}
K={\sum}_{i}[\Tr({\bar\P }_{i}e^{V_N}\P _{i}e^{-V_{M_{i}}})]-c_N\Tr\,V_N-{\sum}_{i}c_{M_i}\Tr(V_{M_{i}})+{\mathcal{L}}(extra)~.
\ee

\begin{figure}[htbp]
\begin{center}
\includegraphics[width=55mm]{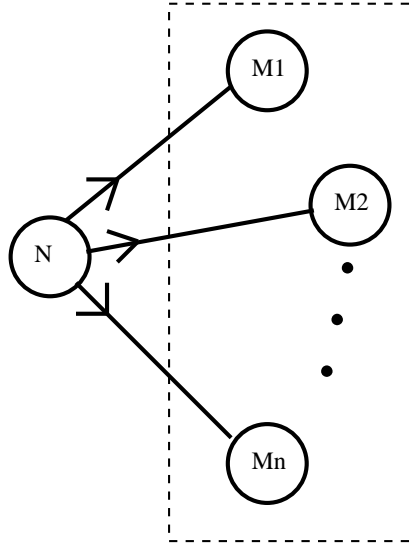}
\caption{One node of a general quiver connecting to many nodes}
\end{center}
\label{fig3}
\end{figure}
\noindent Defining the $({\sum}_{i}M_{i})\!\times\! ({\sum}_{i}M_{i})$ block-diagonal matrix
\be
\label{Vdot}
V_{M}= \left( \begin{array}{ccc} V_{M_{1}} & \dots & 0  \\ \vdots & \ddots & \vdots  \\  0 & \dots & V_{M_n} \end{array} \right) ~,
\ee
and the $N\!\times\! {\sum}_{i}M_{i}$ matrix
\be
\P = (\P_{1} {\dots} \P_{n})~,
\ee
we can rewrite the K\"ahler potential (\ref{kp}) as
\be
K=\Tr({\bar\P }e^{V_N}\P e^{-V_{M}})-c_N\Tr\,V_N-{\sum}_{i}c_{M_i}\Tr(V_{M_{i}})+{\mathcal{L}}(extra)~,
\ee
with $M={\sum}_{i}M_{i}$.
A similar rewriting applies to the dual model, which is now given by
\be
{\tilde{K}}=\Tr({\tP }e^{-V_{M\!-\!N}}{\bar{\tP }}e^{V_{M}})-{\tilde{c}}_{M\!-\!N}\Tr(V_{M\!-\!N})-{\sum}_{i}{\tilde{c}}_{M_i}\Tr(V_{M_{i}})+{\mathcal{L}}(extra)~.
\ee
Note that to prove this generalization of Classical $N=1$ Seiberg duality for the node $N$ we need not consider the gauging of the $M_{i}$ nodes explicitely. Therefore, as far as we are concerned, we can treat them as flavors (the only remnant of their gauge character shows up in the presence of the FI terms, which we do {\it{not}} rewrite using the big flavor matrix $V_{M}$). The advantage of this rewriting is that we can now solve both models exactly as before. The 
equations of motion of $V_N$ and $V_{M\!-\!N}$ imply
\be
\frac{\det(\P e^{-V_{M}}{\bar\P })}{\det({\bar{\tP }}e^{V_{M}}{\tP })}=\det e^{-V_{M}}~,
\ee
which is (\ref{detdet}) with $V_{M}$ given by (\ref{Vdot}). However now $\det e^{-V_{M}}={\prod}_{i} \det e^{-V_{M_{i}}}$ and so compatibility among the FI parameters requires
\be
-c_N&=&{\tilde{c}}_{M\!-\!N}~,\nonumber \\ 
c_{M_i}+c_N&=&{\tilde{c}}_{M_i}~.
\ee
Clearly, when there are $\ell_{\li{NM_i}}$ multiple links between the node $N$ and one or 
more of its neighbors $M_i$, then the same reasoning leads to
\be
-c_N&=&{\tilde{c}}_{M\!-\!N}~,\nonumber \\ 
c_{M_i}+\ell_{\li{NM_i}} c_N&=&{\tilde{c}}_{M_i}~.
\ee

These quivers that we have considered so far are not the most general $N=1$ quivers that one can envisage, as we have assumed that the arrows in the links connecting to the node we are dualizing are all oriented in the same direction. When one tries to dualize an $N=1$ quiver at a node with mixed arrows, one encounters difficulties. 

\begin{figure}[htbp]
\label{inout}
\begin{center}
\includegraphics[width=40mm]{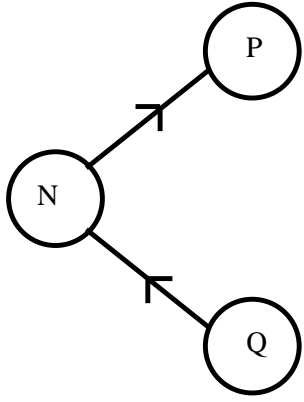}
\caption{Quiver with incoming and outgoing links}
\end{center}
\end{figure}
\noindent We can write the gauged K\"ahler potential for the general quiver in Fig.\ 4 as
\be
K&=&\Tr({\bar\P }_{1}e^{V_N}\P _{1}e^{-V_{P}})+\Tr(\P _{2}e^{-V_N}{\bar\P }_{2}e^{V_{Q}})-c_N\Tr\,V_N\nonumber \\ 
&&~\qquad\qquad -~c_{P}Tr(V_{P})-c_Q Tr(V_Q)+{\mathcal{L}}(extra)~.
\ee
The equations of motion for the gauge field $V_N$ then read
\be
{(\P _{1}e^{-V_P}{\bar\P }_{1})^a}_{b}{(e^{V_N})^b}_c-{(e^{-V_N})^a}_{b}{(\bar\P _{2}e^{V_Q}\P _{2})^b}_c=c_N\,{{\delta}^a}_c~.
\ee
Following our previous discussion it would be natural to conjecture that this quiver is dual to the one with K\"ahler potential 
\be
\tilde{K}&=&\Tr({\tP }_{1}e^{-V_{P\!+\!Q\!-\!N}}{\bar{\tP }}_{1}e^{V_P})+Tr(\bar{\tP }_{2}e^{V_{P\!+\!Q\!-\!N}}\tP _{2}e^{-V_Q})-
\tilde{c}_{P\!+\!Q\!-\!N}\Tr\,V_{P\!+\!Q\!-\!N}\nonumber \\ 
&&\qquad\qquad-~\tilde{c}_{P}\Tr(V_{P})-\tilde{c}_Q\Tr(V_Q)+{\mathcal{L}}(extra)~,
\ee
for which the equations of motion are
\be
{({\tP }_{2}e^{-V_Q}{\bar{\tP }}_{2})^a}_{b}{(e^{V_{P\!+\!Q\!-\!N}})^b}_c-
{(e^{-V_{P+Q-N}})^a}_{b}{(\bar{\tP }_{1}e^{V_{P}}\tP _{1})^b}_c=
c_{P\!+\!Q\!-\!N}\,{{\delta}^a}_c~,
\ee
The respective solutions are
\be
K&=&c_N\left[\ln \det\left(c_N+\sqrt{c_N^{2}+4M_+M_-}\right)+\ln \det(M_+)\right]
\nonumber \\ &&\qquad\qquad-~c_{P}Tr(V_{P})-c_{Q}Tr(V_{Q})+{\mathcal{L}}(extra)~,
\nonumber \\ 
{\tilde{K}}&=&\tilde{c}_{P+Q-N}\left[\ln \det\left(\tilde{c}_{P+Q-N}+\sqrt{{\tilde{c}_{P+Q-N}}^{2}+4{\tilde{M}}_+{\tilde{M}}_-}\right)+\ln \det({\tilde{M}}_+)\right]\nonumber \\ &&\qquad\qquad-~c_{P}Tr(V_P)-c_Q~Tr(V_Q)+{\mathcal{L}}(extra)~,
\ee
where we have defined $M_+=\P _1e^{-V_P}{\bar\P }_1$, $M_-={\bar\P}_2 e^{V_Q}\P_2$, and ${\tilde M}_+={\tP }_{2}e^{-V_Q}{\bar{\tP }}_2$, ${\tilde{M}}_-={\bar{\tP }}_1e^{V_{P}}{\tP }_1$. It is clear that now, because of the presence of the additional terms, the mapping between these two expressions is far from obvious. Inspired by the $N=1$ derivation of $N=2$ Seiberg duality given at the end of the next section, we hope to be able to find a prescription for general $N=1$ quivers by adding appropriate superpotentials. Note also that the original Seiberg duality for this quiver would hold for $P=Q\equiv M$ and would take $N\rightarrow M-N$ plus an additional link connecting the two exterior nodes corresponding to the meson field in the $\bar M \times M$ representation and a superpotential constraint.

\section{$N=2$: The hyperk\"ahler quotient}
\setcounter{equation}{0}
\subsection{Introduction}
We begin by working in $N=2$ projective superspace \cite{pr}. This has the enormous 
advantage of reducing the $N=2$ calculation to the $N=1$ calculation of the previous section. Before embarking on a review of the basics of projective superspace, we summarize the key ideas. The two chiral superfields of the hypermultiplet are encoded in a single polar projective superfield $\u\ofz $; as far as the calculations here are concerned, this behaves exactly as chiral superfield $\P$ in the $N=1$ case. The $N=2$ gauge multiplet is described by a real equatorial projective superfield $V\ofz $, which behaves exactly as the $N=1$ gauge multiplet $V$ of the previous section. Though in this description, it appears that the links once again must have an orientation, there is a another kind of  ``duality'' that reverses the orientation of a link without changing the hyperk\"ahler quotient, and hence the orientation is meaningless. However, to perform the Seiberg-duality, we must again choose the apparent orientations of the links at the node that we are dualizing to be all the same. Finally, in projective superspace, the triplet of FI parameters $b,c,\bar b$ are encoded in a single $N=2$ parameter, which we write as $c\ofz$.

A major difference between these results and the results of the previous section is that whereas the $N=1$ duality is purely classical, because the couplings to hypermultiplets are vector-like and hence nonanomalous, and because of $N=2$ super-nonrenormalization theorems, the $N=2$ Seiberg-duality is a duality relating the vacuum structure of full quantum field theories (in four dimensions) or a full equivalence between conformal field theories (in two dimensions).

As is well known, to carry out the hyperk\" ahler quotient, one has to write the moment map equations for the triholomorphic action of the gauge group on the hypermultiplets; the maps split into a set of holomorphic moment maps (and their conjugates) ($F$-terms), and real moment maps ($D$-terms). To find the quotient space, one restricts to the submanifold given by a level set of the moment maps and divides by the action of the gauge group. Alternatively, the quotient space is found as the set of stable orbits under the complexified gauge group \cite{hklr} subject to the holomorphic moment map constraints.

In the projective superspace formalism, the hyperk\" ahler quotient looks like an ordinary K\"ahler quotient; there are no separate $F$-term constraints, as they  are incorporated in the structure of the supermultiplet. We now give a brief summary of the relevant ideas of projective superspace \cite{pr}.
\subsection{Review of Projective Superspace}
$N=2$ superspace in, \eg, four dimensions, has two sets of spinor derivatives $D_\al^a,\bar D_{a\ad}$, obeying $\{D_\al^a,D_\ba^b\}=\{\bar D_{a\ad},\bar D_{b\bd}\}=0$ and $\{D_\al^a,\bar D_{b\ad}\}=i\de^a_b\pa_{\al\ad}$.  We can find a maximal set of mutually anticommuting derivatives parametrized by a sphere; if we describe the sphere as $\mathbb{C}{\rm P}^1$, then we can write 
\be
\label{prd}
\na_\al\ofz \equiv D_\al^1+\zeta D_\al^2~,\qquad
\bar\na_\ad\ofz \equiv \bar D_{2\ad}-\zeta \bar D_{1\ad}~,
\ee
where $\zeta$ is the usual inhomogenous complex coordinate on $\mathbb{C}{\rm P}^1$. Note that projectively, these derivatives close under an involution $R$ given by composing complex conjugation with the antipodal map on the sphere: $\bar\na_\ad\ofz =(-\zeta) [\na_\al(-1/\bar\zeta)]^\dag$. The basic objects that we consider are {\em projective} superfields $\Psi$ that are annihilated by all the derivatives (\ref{prd}):
\be
\label{prc}
\na_\al\ofz \Psi = \bar\na_\ad\ofz \Psi = 0~.
\ee
Hypermultiplets can be described by arctic superfields $\u\ofz $ that are regular near the north pole of the sphere ($\zeta=0$) and their $R$-conjugates $\bar\u(-1/\zeta)$ that are regular at the south pole ($1/\zeta=0$):
\be
\u=\sum_{i=0}^\infty \u_i\zeta^i~,\qquad\bar\u=\sum_{i=0}^\infty\bar\u_i\inz^i~.
\ee
The constraint (\ref{prc}) implies that the $N=1$ projections of the coefficient superfields $\u_i$ are constrained\footnote{In the literature, some papers use these conventions, and others the complex conjugate conventions.}:
\be
\label{uc}
D^1_\al \u_0 =0~,~~~(D^1)^2\u_1=0~,
\ee
that is, $\u_0$ projects to an antichiral $N=1$ superfield $\bar\P$, $\u_1$ projects to a complex antilinear $N=1$ superfield $\bar\Sigma$, and all the remaining $\u_i$ project to complex unconstrained $N=1$ superfields $X_i$. In this language, the free hypermultiplet Lagrange density is
 \be
 \label{freeu}
 L_{free}=(D^1)^2(\bar D_1)^2 \oint_C\frac{d\zeta}{2\pi\zeta}\, \bar\u\inz\u\ofz~,
 \ee
which is $N=2$ supersymmetric despite the appearance of the explicit $N=1$ spinor derivatives $D^1,\bar D_1$ because the superfields obey the constraint (\ref{prc}). Evaluating the contour integral and projecting to $N=1$ superspace, one finds:
\be
L_{free}=(D^1)^2(\bar D_1)^2 \Big(\bar\P \P-\bar\Sigma\Sigma+\sum_{i=2}^{\infty} (-1)^iX_i\bar X_i\Big)~;
\ee
integrating out the unconstrained fields $X_i$ and dualizing the complex linear superfield $\Sigma$ to a chiral superfield, we obtain the usual Kahler potential for $\mathbb{C}^2$.

Analytic functions of polar multiplets are again polar, and hence it is natural to consider gauge transformations
\be
\label{gtu}
\u'\ofz=e^{i\la\ofz}\u\ofz~~\Rightarrow~~\bar\u'\inz=\bar\u\inz 
e^{-i\bar\la(\frac{-1}\zeta)}~,
\ee
where $\la\ofz$ is an arctic gauge parameter, and its conjugate $\bar\la\inz$ is antarctic.
Clearly, these transformations do not preserve the free Lagrangian (\ref{freeu}); just as in $N=1$ superspace, we introduce a field $V(\zeta,-1/\zeta)$ that ``converts'' $\bar\la$ to $\la$:
\be
\label{gtv}
\Big(e^V\Big)'= \Big(e^{i\bar\la(\frac{-1}\zeta)}\Big)\Big(e^V\Big)\Big(e^{i\la\ofz}\Big)~;
\ee
$V$ is hermitian with respect to the involution $R$, which implies
\be
V_{-i}=(-1)^i\bar V_i~,
\ee
and hence $V$ has singularities at both poles; we call it a tropical multiplet \cite{pr}. For a $U(1)$ subgroup, the gauge transformation on $V\ofz$ is simply
\be\label{gtvu1}
V'=V+i\Big[\bar\la\inz-\la\ofz\Big]~.
\ee
Thus the gauged $N=2$ superspace Lagrange density for an $N=2$ quiver including a piece as in Fig.\ 2 can be written as
\be\label{LgaugeN2}
L&\!=\!&(D^1)^2(\bar D_1)^2 \oint_c \frac{d\zeta}{2\pi\zeta}\Big(\Tr({\bar\u }e^{-V_N}\u e^{V_{N\!+\!M}})-c_N\ofz\, \Tr\,V_N\nonumber\\
&&\qquad\qquad\qquad\qquad\qquad-~c_{N\!+\!M}\ofz \, \Tr\,V_{N\!+\!M}+{\mathcal{L}}(extra)\Big)~.
\ee
Gauge invariance of the $N=2$ FI terms is guaranteed by the constraints on polar multiplets (\ref{uc}) when
\be
c\ofz=\frac{\bar b}\zeta + c - b\zeta~,
\ee
where $c$ is a real constant, because the contour integral in (\ref{LgaugeN2}) picks out only the constrained $\la_0,\la_1,\bar\la_0$, and $\bar\la_1$ parts of (\ref{gtvu1}). Note the close resemblance to the $N=1$ case.

The proof of duality is now identical to the calculation of the previous section. It is nevertheless interesting (and non-trivial) to give the proof directly in $N=1$ superspace. Before performing this calculation, we review how to descend from $N=2$ projective superspace to $N=1$ superspace \cite{pr}. We begin
by factorizing the projective equatorial gauge field $e^{V\ofz}$ into polar but nonprojective factors:
\be\label{Vfact}
e^{V\!\big(\zeta,\frac{-1}\zeta\!\!\big)}\equiv 
e^{V_-\!\big(\!\frac{-1}\zeta\!\!\big)}\,e^{V_+\ofz}~,
\ee
where $V_-$ is the $R$ conjugate of $V_+$. Though $V\ofz$ obeys
$\na\ofz V=0$, the polar factors $V_\pm$ do not; we use them to define
gauge covariant derivatives
\be\label{covdef}
\cd \equiv e^{V_+\ofz}\na e^{-V_+\ofz} =
 -\left(\na e^{-V_-\!\big(\!\frac{-1}\zeta\!\!\big)}\right)
 e^{V_-\!\big(\!\frac{-1}\zeta\!\!\big)}~;
 \ee
comparing the $\zeta$ dependence of the two expressions, we find
\be\label{covd}
\cd_\al\ofz=\cd_\al^1+\zeta\cd_\al^2~,\qquad
\bcd_\ad\ofz=\bcd_{2\ad}-\zeta\bcd_{1\ad}~,
\ee
where $\cd^a_\al,\bcd_{a\ad}$ are the usual ($\zeta$-independent) $N=2$ gauge covariant derivatives.
We also define gauge-covariantly projective hypermultiplet fields
\be
\hat\u\equiv e^{V_+\ofz}\u~,\qquad 
\hat{\bar\u}\equiv \bar\u e^{V_-\!\big(\!\frac{-1}\zeta\!\!\big)}~,
\ee
\be\label{covproj}
\cd\hat\u=\cd\hat{\bar\u}=0~.
\ee
Then $\bar\u e^V \u\equiv\hat{\bar\u}\hat\u$, and (\ref{LgaugeN2}) can be rewritten as:
\be\label{LhatN2}
L&\!=\!&(D^1)^2(\bar D_1)^2 \oint_c \frac{d\zeta}{2\pi\zeta}\Big(\Tr(\hat{\bar\u }\hat\u) -c_N\ofz\, 
\Tr\,V_N-c_{N\!+\!M}\ofz \, \Tr\,V_{N\!+\!M}+{\mathcal{L}}(extra)\Big)~.\nonumber\\
\ee
Furthermore, the constraints (\ref{covproj}) imply
\be
\cd^1_\al\hat\u_0=0~,\qquad
(\cd^1)^2\hat\u_1=\frac12\{\cd^{1\al},\cd^2_\al\}\hat\u_0
\equiv \bar W\u_0~,
\ee
where $\bar W$ is the $N=2$ superfield strength and reduces to an $N=1$ (anti)chiral superfield $\hat{\bar S}$; thus $\hat\u_0$ reduces to an $N=1$ covariantly (anti)chiral superfield $\hat{\bar\P}_+$, $\hat\u_1$ reduces to an $N=1$ modified complex (anti)linear superfield $\hat{\bar\Sigma}$ obeying $\cd^2\hat{\bar\Sigma}=\hat{\bar S}\hat{\bar\P}_+$, and the remaining $\hat\u_i$ reduce to $N=1$ unconstrained superfields. We can now perform the $\zeta$ integral in (\ref{LhatN2}) and eliminate the auxiliary superfields to find
\be
L&\!=\!&(D^1)^2(\bar D_1)^2[\Tr(\hat{\bar\P}_+\hat\P_+ -\hat{\bar\Sigma}\hat\Sigma 
-c_N\,\Tr\,V_N-c_{N\!+\!M}\,\Tr\,V_{N\!+\!M}]\nonumber\\
&&~~
-\Big(\!(D^1)^2 [b_N \, \Tr\,S_N+b_{N\!+\!M} \, \Tr\,S_{N\!+\!M}] + {\mathrm{h.c.}}\!\Big)+ {\mathcal{L}}(extra)~.
\ee
To obtain the final $N=1$ Lagrange density, we impose the chiral constraint
$\bcd^2\hat\Sigma=\hat\P_+ \hat S$ by a covariantly chiral Lagrange multiplier $\hat\P_-$, and integrate out $\hat\Sigma$; finally, we introduce
the $N=1$ gauge multiplet $V$ and write the $N=1$ action in terms of $N=1$ chiral superfields $\P_\pm,S$:
\be\label{N1N2}
L&\!=\!&(D^1)^2(\bar D_1)^2
[\Tr({\bar\P_+}e^{V_N}\P_+ e^{-V_{N\!+\!M}})+
\Tr({\P_-}e^{-V_N}\bar\P_- e^{V_{N\!+\!M}})
\nonumber\\
&&\qquad\qquad\qquad\qquad-~c_N\Tr\,V_N-
c_{N\!+\!M}\Tr\,V_{N\!+\!M}]
\nonumber\\
&&~~+\Big(\!(D^1)^2 [\Tr(\P_-S_N\P_+)-\Tr(\P_+S_{N\!+\!M}\P_-) \nonumber\\
&&\qquad\qquad\qquad \qquad
-~b_N \, \Tr\,S_N-b_{N\!+\!M} \, \Tr\,S_{N\!+\!M}]+
{\mathrm{h.c.}}\!\Big)+ {\mathcal{L}}(extra)~.
\ee
\subsection{$N=1$ calculation for $N=2$ Seiberg duality}
We now turn to the $N=1$ superspace proof of $N=2$ Seiberg duality.
We first consider the special case when the $N=2$ FI terms $b_N$ vanish, and 
subsequently the general case (which involves some extra matrix identities).

Varying the kinetic terms in (\ref{N1N2}) with respect to $V_N$ gives
the D-flatness equations:
\be
\label{dflatness}
M_+~e^{V_N}~-~e^{-V_N}M_-=c_N{\bf 1}_N~,
\ee
with $M_+ = \P_+e^{-V_{N\!+\!M}}\bar\P_+$ and $M_- = \bar\P_- e^{V_{N\!+\!M}}\P_-$. The F-flatness equations read
\be
\label{fflatness}
\P_+\P_- =0~
\ee
where we use $b_N=0$. Using the $U(N)$ gauge symmetry $\P_+$ can be chosen to have the form
\be
\label{gaugechoice}
\P_+ =
\left(
\begin{array}{cc}
{\bf 1}_{N\!\times\! N} & Q_{N\!\times\! M}
\end{array}
\right)~.
\ee
Equation (\ref{fflatness}) then determines $\P_-$ to be
\be
\P_- =
\left(
\begin{array}{c}
-QB \\
B
\end{array}
\right)~,
\ee
where $B$ is an arbitrary $M\!\times\! N$ matrix. Finally, writing the solution to (\ref{dflatness}) in the form 
\be
e^{V_N} &=&{1\over 2} M_+^{-1}\left(c_N + \sqrt{c_N^2 + 4M_+ M_-}\right) ~,
\ee
we can write the relevant part of the quotient hyperk\"ahler potential as
\be
\label{dkp}
K&=&\Tr\left[\sqrt{c_N^2 +4M_+ M_-}\right]-c_N\,\Tr\ln\left(c_N+\sqrt{c_N^2+4M_+ M_-}\right) \nonumber\\
&&\qquad\qquad\qquad~+~c_N\,\Tr\ln M_+  -c_{N\!+\!M}\,\Tr\, V_{N\!+\!M}+{\cal L}(extra) ~,
\ee
with $M_\pm$ (defined below (\ref{dflatness})) constrained by (\ref{fflatness}). 

The same hyperk\"ahler potential (\ref{dkp}) can be arrived at starting from
\be
\tilde K &=&(D^1)^2 \ (\bar D_1)^2 [ \Tr (\tP_+ e^{-V_M} \bar{\tP}_+ e^{V_{N\!+\!M}})+ \Tr(\bar{\tP}_- e^{V_M}\tP_- e^{-V_{N\!+\!M}})\nonumber \\
&&\qquad\qquad\qquad-\, \tilde c_M\Tr\,V_M  - 
\tilde c_{N\!+\!M}\Tr\,V_{N\!+\!M} ]\nonumber \\
&&~~+\Big(\!(D^1)^2 [\Tr(\tP_+S_M\tP_-)-\Tr(\tP_-S_{N\!+\!M}\tP_+)]+
{\mathrm{h.c.}}\!\Big)+ {\mathcal{L}}(extra)~.
\ee
Choosing the $U(M)$ gauge 
\be\label{gcm}
{\tP}_+ =\left(
\begin{array}{c} -Q_{N\!\times\! M} \\{\bf 1}_{M\!\times\! M}\end{array}\right)~,
\ee
the F-term constraints ${\tP}_- {\tP}_+ = 0$ give 
\be
{\tP}_-=\left(
\begin{array}{cc} C_{M\!\times\! N} & (CQ)_{M\!\times\! M}
\end{array}\right)~,
\ee
where the matrix $C$ parameterizes the coordinates of the quotient. Further solving the D-flatness equations
\be
e^{V_M}\tilde M_- - \tilde M_+ e^{-V_M} = \tilde c_M {\bf 1}_{M\!\times\! M}~,
\ee
where $\tilde M_- =\tP_- e^{-V_{N\!+\!M}}\bar{\tP}_-$ and $M_+=\bar{\tP}_+e^{V_{N\!+\!M}}\tP_+$ as
\be
e^{-V_M} = {1\over 2} \tilde M_+^{-1} \left(-\tilde c_M + \sqrt{ \tilde c_M^2 + \tilde M_+ \tilde M_-} \right)~,
\ee
gives the hyperk\"ahler potential 
\be
\label{nkp}
\tilde K&=&\Tr \left[\sqrt{\tilde c_M^2 +4 \tilde M_+ \tilde M_-}\right] + \tilde c_M\Tr\ln\left(-\tilde c_M + \sqrt{\tilde c_M^2+4\tilde M_+ \tilde M_-}\right) \nonumber\\
&&\qquad\qquad -~\tilde c_M\Tr\ln \tilde M_+ -\tilde c_{N\!+\!M}\Tr\,V_{N\!+\!M}+{\cal L}(extra) ~.
\ee
Noting that the respective solutions to the F-flatness equations give $\P_- \P_+ = \tP_+\tP_-$ after setting $C=B$; thus one obtains $\Tr f( M_+ M_-) =\Tr f(\tilde M_+ \tilde M_-)$ for any function $f$, which tells us that the first two terms in (\ref{dkp}) and (\ref{nkp}) are identical provided $c_N=-\tilde c_M$. The rest of the proof then reduces to the argument given in the k\"ahler case, which implies the map among the rest of the FI parameters, $c_{N\!+\!M}+c_N =\tilde c_{N\!+\!M}$.

We now turn to the general situation when all the $N=2$ FI-terms are nonvanishing. In this case, the holomorphic constraints (F-term equations) are modified and the relation between $M_+M_-$ and $\tilde M_+\tilde M_-$ is more subtle. The D-term equations remain unaltered. The direct hyperk\"ahler potential is therefore still given by (\ref{dkp})
\be
\label{dkpbis}
K&=&\Tr\left[\sqrt{c_N^2 +4M_+ M_-}\right]-c_N\,\Tr\ln\left(c_N+\sqrt{c_N^2+4M_+ M_-}\right) \nonumber\\
&&\qquad\qquad\qquad~+~c_N\,\Tr\ln M_+  -c_{N\!+\!M}\,\Tr\, V_{N\!+\!M}+{\cal L}(extra) ~,
\ee
but with $M_+ = \P_+e^{-V_{N\!+\!M}}\bar\P_+$ and $M_- = \bar\P_- e^{V_{N\!+\!M}}\P_-$ subject to
\be
\label{fflatnessbis}
\P_+\P_- =b_N {\bf 1}_{N\!\times\! N}~.
\ee
Similarly, the dual hyperk\"ahler potential is given by (\ref{nkp})
\be 
\label{nkpbis}
\tilde K&=&\Tr \left[\sqrt{\tilde c_M^2 +4 \tilde M_+ \tilde M_-}\right] + \tilde c_M\Tr\ln\left(-\tilde c_M + \sqrt{\tilde c_M^2+4\tilde M_+ \tilde M_-}\right) \nonumber\\
&&\qquad\qquad -~\tilde c_M\Tr\ln \tilde M_+ -\tilde c_{N\!+\!M}\Tr\,V_{N\!+\!M}+{\cal L}(extra) ~,
\ee 
with $\tilde M_- =\tP_- e^{-V_{N\!+\!M}}\bar{\tP}_-$ and $M_+=\bar{\tP}_+e^{V_{N\!+\!M}}\tP_+$ now subject to
\be
\label{ahfflatnessbis}
\tP_- \tP_+ = \tilde b_M {\bf 1}_{M\!\times\! M}~.
\ee
One can still make the same $U(N)$ and $U(M)$ gauge choices (\ref{gaugechoice},\ref{gcm}) as in the $b_N=0$ case
\be
\label{bnzchoices}
\P_+ =
\left(
\begin{array}{cc}
{\bf 1}_{N\!\times\! N} & Q_{N\!\times\! M}
\end{array}
\right)~,\qquad\qquad\tP_+ =\left(\begin{array}{c} -Q_{N\!\times\! M} \\ {\bf 1}_{M\!\times\! M}\end{array}\right)~.
\ee
Solving the constraints (\ref{fflatnessbis}) and (\ref{ahfflatnessbis}) we obtain,
\be
\P_-=
\left(
\begin{array}{c}
b_N{\bf 1}-QB \\
B \end{array}
\right)~,\qquad\qquad\tP_-=\left(
\begin{array}{cc} C & \tilde b_M {\bf 1} + CQ
\end{array}\right)~.
\ee
As in the $b_N=0$ case, we choose $C=B$; this gives
\be
\label{mesonsbis}
\P_-\P_+&=&
\left(
\begin{array}{cc}
b_N{\bf 1}-QB & b_N Q-QBQ \\
B & BQ
\end{array}
\right)\nonumber\\
\tP_+ \tP_- &=&
\left(
\begin{array}{cc}
-QB & -\tilde b_M Q - QBQ \\
B & \tilde b_M {\bf 1} + BQ
\end{array}
\right)~.
\ee
We first prove that the first two terms in (\ref{dkpbis}) and (\ref{nkpbis}) are equal, up to irrelevant constant terms. To see this we first note that for $b_N , \tilde b_M\neq 0$, we can factorize $\P_-\P_+$ and $\tP_+\tP_-$ as
\be
\label{rewrite}
\P_-\P_+&=&
b_N \left(
\begin{array}{cc}
{\bf 1}-{1\over b_N}QB & -Q \\ {1\over b_N}B & {\bf 1}
\end{array}\right)
\left(
\begin{array}{cc}
{\bf 1}_{N\!\times\! N} & 0 \\ 0 & 0
\end{array}
\right)
\left(
\begin{array}{cc}
{\bf 1} & Q \\ -{1\over b_N} B & {\bf 1}-{1\over b_N}BQ
\end{array}
\right)\nonumber \\
\tP_+\tP_- &=&
\tilde b_M \left(
\begin{array}{cc}
{\bf 1}+{1\over \tilde b_M}QB & -Q \\ -{1\over \tilde b_M}B & {\bf 1}
\end{array}
\right)
\left(
\begin{array}{cc}
0 & 0 \\ 0 & {\bf 1}_{M\!\times\! M}
\end{array}
\right)
\left(
\begin{array}{cc}
{\bf 1} & Q \\ {1\over \tilde b_M}B & {\bf 1}+{1\over \tilde b_M}BQ
\end{array}
\right)~.
\ee
However,
\be
\label{unity}
\left(
\begin{array}{cc}
{\bf 1}-{1\over b_N}QB & -Q \\ {1\over b_N}B & {\bf 1}
\end{array}\right)
\left(
\begin{array}{cc}
{\bf 1} & Q \\ -{1\over b_N}B & {\bf 1}-{1\over b_N}BQ
\end{array}
\right)=
\left(
\begin{array}{cc}
{\bf 1} & 0 \\
0 & {\bf 1}
\end{array}
\right) ~;
\ee
hence, if we set $b_N = -\tilde b_M$, we can write
\be
\label{MMNN}
\P_-\P_+=b_N\, e^{i\Lambda}P_N e^{-i\Lambda}~,~~
\tP_+\tP_-=b_N\, e^{i\Lambda}P_M e^{-i\Lambda}~,~~
e^{i\Lambda}\equiv \left(
\begin{array}{cc}
{\bf 1}-{1\over b_N}QB & -Q \\ {1\over b_N}B & {\bf 1}
\end{array}\right)~,
\ee
and $P_N,P_M$ are the diagonal projectors in (\ref{rewrite}). Thus, for any function $f$,
\be
\label{M+M-}
\Tr f(M_+M_-)&=&  \Tr \,f\! \left(|b_N|^2 P_N \left(e^{-i\Lambda} e^{-V_{N\!+\!M}}e^{i\bar\Lambda}\right) P_N \left(e^{-i\bar\Lambda}e^{V_{N\!+\!M}}e^{i\Lambda}\right)\right)~,\nonumber \\
\Tr f(\tilde M_+ \tilde M_-)&=&\Tr\,f\!\left( |b_N|^2 P_M \left(e^{-i\Lambda} e^{-V_{N\!+\!M}}e^{i\bar\Lambda}\right) P_M \left(e^{-i\bar\Lambda}e^{V_{N\!+\!M}}e^{i\Lambda}\right)\right)~.
\ee 
We now use the identity
\be
\label{crident}
\Tr\, f\! \left( P_N \mO P_N \mO^{-1}\right)& =& \Tr \,f\!\left(((1-P_M)\mO(1- P_M) \mO^{-1}\right) \nonumber\\ 
&=&\Tr \,f\!\left(P_M\mO P_M \mO^{-1}\right)+\Tr\,f(1-P_M)-\Tr\,f(P_M)\nonumber\\ 
&=&\Tr \,f\!\left(P_M\mO P_M \mO^{-1}\right)+\Tr\,f(P_N)-\Tr\,f(P_M)\nonumber\\ 
&=&\Tr \,f\!\left(P_M\mO P_M \mO^{-1}\right)+(N\!-\!M)(f(1)\!-\!f(0))~,
\ee
where $\mO$ is any invertible matrix.

Expressing now the first two terms in the hyperk\"ahler potentials (\ref{dkpbis}) and (\ref{nkpbis}) in terms of (\ref{M+M-}), and applying (\ref{crident}), we see that modulo irrelevant constant terms these are equal provided $c_N =-\tilde c_M$. Further, equality of the third terms now follows from the argument given in the K\"ahler case, which again gives $c_{N\!+\!M}+c_N =\tilde c_{N\!+\!M}$.

To finish the proof of the duality, one only needs to see how the rest of the complex FI parameters transform. Writing the holomorphic moment map constraints at nodes $N$ and $N\!+\!M$ gives
\be
\label{nodeN}
\P_+\P_-&=&b_N{\bf 1}_{N\!\times\!N}\nonumber\\
-\P_-\P_+ +{\rm extra}&=&
b_{N\!+\!M}{\bf 1}_{(N\!+\!M)\!\times\!(N\!+\!M)}~.
\ee
Similarly, at the dual nodes $M$ and $N\!+\!M$
\be
\label{nodeM}
\tP_-\tP_+&=&\tilde b_M {\bf 1}_{M\!\times\!M}\nonumber\\
-\tP_+\tP_- +{\rm extra}&=&\tilde b_{N\!+\!M}{\bf 1}_{(N\!+\!M)\!\times\!(N\!+\!M)}~,
\ee
where ``extra'' is the contribution coming from the additional arrows connected to the $N\!+\!M$ node, and therefore is the same in (\ref{nodeN}) and (\ref{nodeM}). Combining the traces of (\ref{nodeN}) and (\ref{nodeM}), and taking into account $\tilde b_M = -b_N$ we obtain
\be
\tilde b_{N\!+\!M}=b_{N\!+\!M}+b_N~.
\ee
To summarize, we have seen that, as expected by symmetry under $SU(2)_R$ rotations, the triplet of FI parameters transforms as
\be
\label{fitraf}
(c_N,b_N,\bar b_N)&\rightarrow& -(c_N,b_N,\bar b_N)~,\nonumber\\
(c_{N\!+\!M},b_{N\!+\!M},\bar b_{N\!+\!M})&\rightarrow&(c_{N\!+\!M}+c_N,b_{N\!+\!M}+b_N,\bar b_{N\!+\!M}+\bar b_N)~.
\ee
\section{Relation to other approaches}
\setcounter{equation}{4}
The $N=2$ Seiberg duality we have explored in this paper was already noticed in \cite{ckv}, where, in the particular case where the quiver corresponds to the Dynkin diagram of an $ADE$ group, it was interpreted as Weyl reflections around primitive roots. However in this paper we have shown by direct computation that the quiver duality is in some sense more fundamental, as it can be seen as an algebraic fact that holds for any quiver and any representation thereof.

To make the relation with the $ADE$ case more explicit, recall that the adjacency matrix of a quiver $A_{ij}\equiv$ {\em the number of links between nodes $i$ and $j$} is closely related to the Cartan matrix $C_{ij}=e_i {\cdot} e_j$ of a Lie group, where the $e_i$ are the simple roots and appear as the nodes in the quiver diagram. The precise relationship is
\be
\label{avsc}
C_{ij}=2\delta_{ij}-A_{ij}~.
\ee
Weyl reflections act on simple roots  as
\be
\label{weylgroupaction}
 e_j -C_{ji}e_i \equiv {\cal W}_{ij}e_j~.
\ee
Now, for a quiver theory with gauge group $\prod_i^n U(N_i)$ the vector ${\cal N}\equiv (N_1,\ldots,N_n)$ can be seen as a positive root of the algebra associated with the Cartan matrix $C_{ij}$ by writing ${\cal N}=\sum_i^n N_i e_i$. Denoting  the vector transformed under (\ref{weylgroupaction}) by ${\cal N}'=\sum_i^n N'_i e'_i$, and requiring ${\cal N}'={\cal N}$, which was interpreted as a brane charge conservation condition in \cite{ckv}, we obtain
\be
\label{wa}
N_i\rightarrow -N_i - \sum_{j\neq i} C_{ij} N_j,~~~~~~~N_{j\neq i}\rightarrow N_j~.
\ee
The FI parameters are associated to the $U(1)$ centers of each $U(N_i)$, and they transform as simple roots. For Seiberg duality around node $i$ (Weyl reflection around node $i$), using (\ref{weylgroupaction}) we have
\be
b_j \rightarrow b_j - C_{ji} b_i~,
\ee
which is precisely (\ref{fitraf}).

\section{Conclusions}
In this paper we have given an elementary explicit proof of $N=2$ Seiberg duality for general $N=2$ quivers. We have found the mapping between the $FI$ parameters of the dual models. The proof in $N=1$ superspace is quite subtle. The $N=2$ projective superspace approach simplifies the proof and gives a unified picture of $N=2$ quivers and some particular $N=1$ quivers (when all arrows point in a single direction). This implies a {\it classical} duality for $N=1$ {\it K\" ahler} potentials. We would like to be able to generalize this to arbitrary $N=1$ quivers with an appropriate superpotential, and explore the relation to the usual $N=1$ Seiberg duality.  

\vskip 0.5 cm {\bf{Acknowledgments}}\\
The authors are grateful to Warren Siegel and Cumrun Vafa for illuminating suggestions, and Anthony Knapp for supplying the proof in the Appendix. MR is happy to thank Ken Intrilligator for helpful comments. This work was supported in part by NSF Grant No.\ PHY-0098527.

\appendix
\section{Proof of uniqueness of self-dual quivers}
\setcounter{equation}{0}
\begin{figure}[htbp]
\label{Appendixfig}
\begin{center}
\includegraphics[width=115mm]{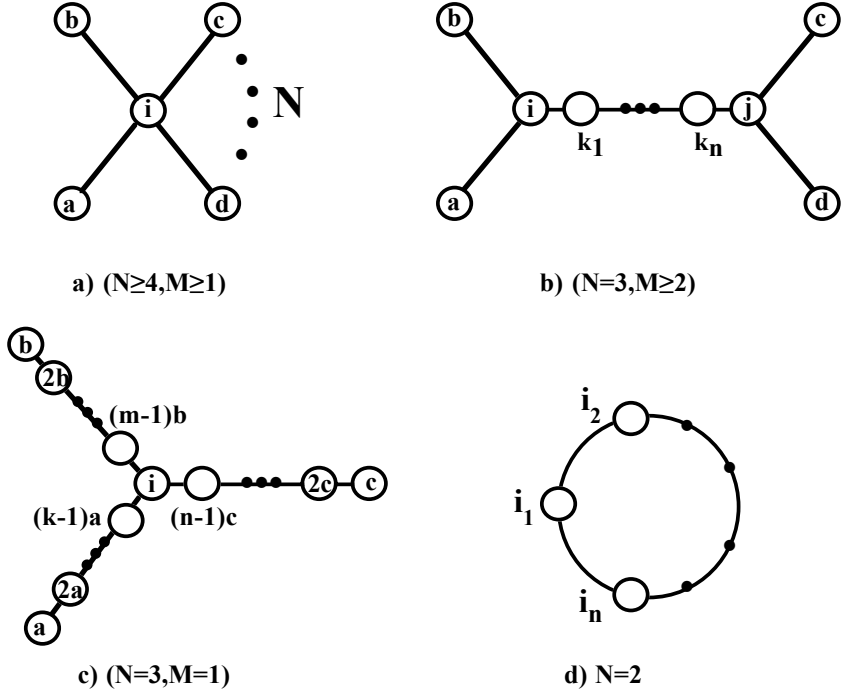}
\caption{Quivers of various types}
\end{center}
\end{figure}

In this appendix we give a proof that extended $\hat A\hat D\hat E$ diagrams are the only labeled quivers that are self-dual under $N=2$ Seiberg duality. This is equivalent to the statement that these are the only superconformal Quiver Theories \cite{ckv,He}.

Consider connected quiver diagrams with nodes $i,j,\ldots,a,b,\ldots,$ with indices $n_i,n_j,\ldots,$ $n_a,n_b,\ldots,$ {\sl etc}.  We introduce the following nomenclature: an $(N,M)$ 
quiver has $M$ nodes with $N$ neighbors, as well as possible nodes with fewer neighbors.

We use the following general statement throughout: if $i$ has $j$ as a neighbor (plus possibly additional neighbors), self-duality at $i$ imposes
\be
2n_{i}{\geq}n_{j}~;
\ee
equality holds if and only if $i$ has only $j$ as a neighbor.\\

We first prove that the only self-dual quiver with $(N{\geq}4,M{\geq}1)$ has $N=4$, $M=1$. Consider a quiver with a node $i$ with four or more neighbors (Fig.~5(a)). 
Self-duality at $i$ requires
\be
\label{aks}
2n_{i}=n_{a}+n_{b}+n_{c}+n_{d}+\ldots
\ee
whereas self-duality at $a,b,c,d,\ldots$ imposes
\be
\label{nsd}
2n_{a}{\geq}n_{i},~~~2n_{b}{\geq}n_{i},~~~2n_{c}{\geq}n_{i},~~~2n_{d}{\geq}n_{i}, ~~~\dots~.
\ee
Adding these up gives $4n_{i}~{\leq}~2n_{a}+2n_{b}+2n_{c}+2n_{d}+\ldots$. Comparing with twice (\ref{aks}) shows that the expression $+\ldots$ in (\ref{aks}) is less or equal than zero, and hence equals zero. Thus $i$ has just $a,b,c,d$ as neighbors, which in turn have no other neighbors. The diagram is the unique self-dual quiver with four nodes with label $n$ connected to a central node with label $2n$, which is the extended Dynkin diagram $\hat D_4$.

With the previous result, we are only left with diagrams whose highest node is a triple node. Consider then the cases $(3,M{\geq}2)$. Pick two of these triple nodes, such that there is no triple node on some path connecting them (Fig.~5(b)). Again, for self-duality
\be
\label{a4}
2n_{i}=n_{a}+n_{b}+n_{k_1},~~~2n_{j}=n_{k_{n}}+n_{c}+n_{d}\nonumber\\
2n_{k_1}=n_{i}+n_{k_2},~~\ldots~~2n_{k_n}=n_{k_{n-1}}+n_{j}
\ee
and
\be
\label{90}
2n_{a}{\geq}n_{i},~~~2n_{b}{\geq}n_{i},~~~2n_{c}{\geq}n_{j},~~~2n_{d}{\geq}n_{j}~.
\ee
Adding all the equations in (\ref{a4}) and multiplying by two one gets $2n_{i}+2n_{j}=2n_{a}+2n_{b}+2n_{c}+2n_{d}$, while (\ref{90}) gives $2n_{a}+2n_{b}+2n_{c}+2n_{d}~{\geq}~2n_{i}+2n_{j}$. Then equality must hold in (\ref{90}), $a,b,c,d$ have only one neighbor and $n_{a}=n_{b}$ and $n_{c}=n_{d}$. This also fixes $2n_{a}=n_{i}=n_{k_{1}}=\ldots=n_{k_{n}}=n_{j}$, which yields $n_b = n_c$. Therefore the quiver corresponds to the extended Dynkin diagram of ${\hat{D}}_{n+5}$.

Consider now diagrams of type $(3,1)$. A series of nodes with exactly two neighbors have indices $n$ that form an arithmetic progression. Two cases are to be considered. If two of the legs emanating from the triple node form a closed loop, self-duality at each of the nodes in the loop implies that these have all the same label. However, for the single triple node self-duality cannot hold, since it is attached to one more nodes whose label is, by assumption, non zero. This rules out the possibility of these configurations. On the other hand, if no two legs close to form a loop we have the quiver depicted in Fig.~5(c).  At the $i$-th node we have the condition
\be
\label{a6}
ka=mb=nc~.
\ee
So $a,b,c$ are rational multiples of something. Without loss of generality we may take them as positive integers. The self-duality condition at the triple node implies
\be
\label{a7}
2ka=(k-1)a+(m-1)b+(n-1)c\\{\nonumber}
2mb=(k-1)a+(m-1)b+(n-1)c\\{\nonumber}
2nc=(k-1)a+(m-1)b+(n-1)c
\ee
Adding the equations in (\ref{a7}) yields $3(a+b+c)=ka+mb+nc$, and this is $3ka=3mb=3nc$ by (\ref{a6}). So $(a+b+c)=ka=mb=nc$. Therefore $a,b,c$ divide $a+b+c$, and in particular any of $a,b,c$ divides the sum of the other two. Assume without loss of generality $a~{\geq}~b~{\geq}~c$. Put $b+c=xa$ ($x$ integer), then $xb~{\leq}~xa=b+c~{\leq}~2b$. Therefore $x$ equals 1 or 2.
 
If $x=2$ then $2a=b+c$, and $a~\geq~b~\geq~c$ therefore forces $a=b=c$. This fixes the quiver to be the extended Dynkin diagram of  ${\hat{E}}_{6}$.

If $x=1$, $a=b+c$, and in particular $b$ divides $a-c$. But $b$ also divides $a+c$, which means that $b$ divides $2a$ and $2c$. Since $c~{\leq}~b$, $2c=yb~{\leq}~2b$ with $y$ integer. So $y$ equals 1 or 2. If $y=2$ then $b=c$ and $a=b+c$, which means $(a,b,c)=(2c,c,c)$. The quiver is then the extended Dynkin diagram of ${\hat{E}}_{7}$. 

If $y=1$ then $b=2c$, with $a=b+c$, which tells us $(a,b,c)=(3c,2c,c)$. This is the extended Dynkin diagram for ${\hat{E}}_{8}$.

Finally, we are only left with diagrams whose highest node has two neighbors (Fig. 5(d)). It is straighforward to check that the only self-dual finite diagrams correspond to ${\hat{A}}_{n}$.

The same kind of arguments may be used to conclude that the only self-dual quiver including multiple links between any nodes is $\hat A_1$.

\end{document}